\documentclass[aps,pre,twocolumn,showpacs]{revtex4}
\usepackage{epsfig}
\usepackage{times}
\usepackage{amsmath}

\usepackage{tipa}
\usepackage{amssymb}
\usepackage{txfonts}
\usepackage{float}
\usepackage[loose,normalsize,hang]{subfigure}

\begin{document}

\title{The robustness of interdependent networks under the interplay between cascading failures and virus propagation}

\author{Dawei Zhao$^{1\ast}$, Zhen Wang$^{2}$,
Gaoxi Xiao$^{3,4}$, Bo Gao$^{5}$, and Lianhai Wang$^{1}$} \affiliation {$^{1}$ Shandong Provincial Key Laboratory of Computer
Networks, Shandong Computer Science Center (National Supercomputer
Center in Jinan), Jinan 250014,
China.\\ $^{2}$ Interdisciplinary Graduate School of Engineering Sciences, Kyushu University, Kasuga-koen, Kasuga-shi, Fukuoka 816-8580, Japan.\\ $^{3}$ School of Electrical and Electronic Engineering, Nanyang Technological University, 50 Nanyang Avenue, Singapore 63979.\\ $^{4}$
Complexity Institute, Nanyang Technological University, Singapore.\\
$^{5}$ School of Computer Information management, Inner Mongolia University of Finance and Economics, Hohhot 010051, China.\\ $^{\ast}${zhaodw@sdas.org}}

\begin{abstract}
Cascading failures and epidemic dynamics, as two successful application realms of network science, are usually investigated separately. How do they affect each other is still one open, interesting problem. In this letter, we couple both processes and put them into the framework of interdependent networks, where each network only supports one dynamical process. Of particular interest, they spontaneously form a feedback loop: virus propagation triggers cascading failures of systems while cascading failures suppress virus
propagation. Especially, there exists crucial threshold of virus transmissibility, above which the interdependent networks collapse completely. In addition, the interdependent networks will be more vulnerable if the network supporting virus propagation has denser connections; otherwise the interdependent systems are robust against the change of connections in other layer(s). This discovery differs from previous framework of cascading failure in interdependent networks, where better robustness usually needs denser connections. Finally, to protect interdependent networks we also propose the control measures based on the identification capability. The larger this capability, more robustness the interdependent networks will be.
\end{abstract}

\pacs{89.75.-k,64.60.aq,05.45.-a} \maketitle

\section{Introduction}

During the past years, complex networks have proven to be a successful tool in describing a large variety of real-world complex systems, ranging from biological, technological, social to information, engineering, and physical systems \cite{newman2010networks,barrat2008dynamical}. The investigations of the structure and dynamics of complex networks have triggered enormous interests, and a lot of remarkable results have been achieved \cite{newman2002random,pastor2015epidemic,cohen2010complex,song2005self,wang2013impact,wang2011coveting,perc2015double,song2006origins}. However, vast majority of existing works mainly focus on single networks that are isolated from each other, despite of the fact that many real-world networks usually interact with and depend on each other. In 2010, Buldyrev et al. \cite{buldyrev2010catastrophic} proposed a new model of networks, so-called interdependent networks, and developed theoretical framework to study the cascading failures of interdependent systems caused by random node removal. Surprisingly, they found that systems made of interdependent networks would be intrinsically more fragile than each isolated component. After that, much research attention moves to more complicated yet more realistic multilayer networks, mainly including interdependent networks \cite{son2012percolation,parshani2011inter,gao2012networks,wang2014self,dong2013robustness}, interconnected networks \cite{saumell2012epidemic,dickison2012epidemics,zhao2013identifying,radicchi2013abrupt,de2014navigability} and multiplex networks \cite{nicosia2013growing,zhao2014immunization,zhao2014multiple,boccaletti2014structure}.

Cascade of failures, as one of the hottest research topics in network science, has attracted great attentions after the seminal idea of Buldyrev \cite{son2012percolation,parshani2011inter,gao2012networks,wang2014self,dong2013robustness,crucitti2004model,wang2007high,huang2011robustness,shao2011cascade,parshani2010interdependent,schneider2013towards}. For example, Parshani et al. explored the influence of degree correlation on the robustness of interdependent networks to cascading failures, and found that the systems become more robust when they share higher inter-similarity \cite{parshani2011inter}. By mapping the random attack to the targeted attack problem, Huang et al. evaluated the cascading failures in interdependent networks under an initial targeted attack \cite{huang2011robustness}. Shao et al. developed a theoretical framework for understanding the cascade process of failures in interdependent networks with a random number of support and dependence relationships \cite{shao2011cascade}. Refs. \cite{parshani2010interdependent,schneider2013towards} showed that when a small fraction of autonomous nodes were properly selected, the nature of the percolation transition changed from discontinuous to continuous fashions and the cascading failures could be largely suppressed.

Epidemic dynamic \cite{pastor2015epidemic}, as another rapidly developing research area in network science, is broadly used to mimic many real propagation processes, such as disease in human contact networks \cite{salathe2010high,wang2014spatial}, information and rumor in social networks \cite{min2014layer}, and virus in computer or communication networks \cite{zhao2013efficient,gao2013modeling}. Understanding the epidemic spreading processes is thus crucial for developing efficient methods to either prevent propagation of disease, rumor and virus, or accelerate information dissemination. At present, the most popular  models to describe the propagation of epidemic include susceptible-infected (SI) model, susceptible-infected-susceptible (SIS) model, and susceptible-infected-recovered (SIR) model \cite{pastor2015epidemic}. Like other dynamic processes upon networks \cite{boccaletti2014structure}, the recent concerns of epidemic spreading also extend from single networks to multilayer networks \cite{saumell2012epidemic,dickison2012epidemics,zhao2014multiple,min2014layer}.

In spite of great progress of recent years, cascading failures and epidemic dynamics are usually considered as two irrelevant research topics and studied separately. However, in many real world systems, cascading failures and epidemic dynamics often influence and interact with each other. For example, the virus propagation on communication network not only causes node failure or load redistribution of communication network, but also triggers the collapse of other related networks like power grid due to the interdependency relationships between them, thus resulting in cascading failures in interdependent systems. The cascade of failure on other networks in turn enables more nodes or fragmentation to be removed in communication network, which thus suppresses the propagation of the virus. In particular, if the virus is not completely suppressed, it will lead to new cash of nodes and thus triggers successive cascading failures in the interdependent networks. Compared with the existing researches on the robustness of interdependent networks, the above case introduces a novel and much more severe attack method for interdependent networks.

Aim to this issue, here we develop a new framework where the virus propagation could induce cascading failures and cascading failures are able to suppress virus propagation (i.e. forming feedback between cascade dynamics and epidemic dynamics). By means of numerous simulations, we will investigate the interplay of both processes in interdependent scale-free (SF) networks \cite{barabasi1999emergence}, and explore the robustness of interdependent networks under this novel setup.

\begin{figure}[!htb]
\includegraphics[scale=1.1,trim=0 0 0 0]{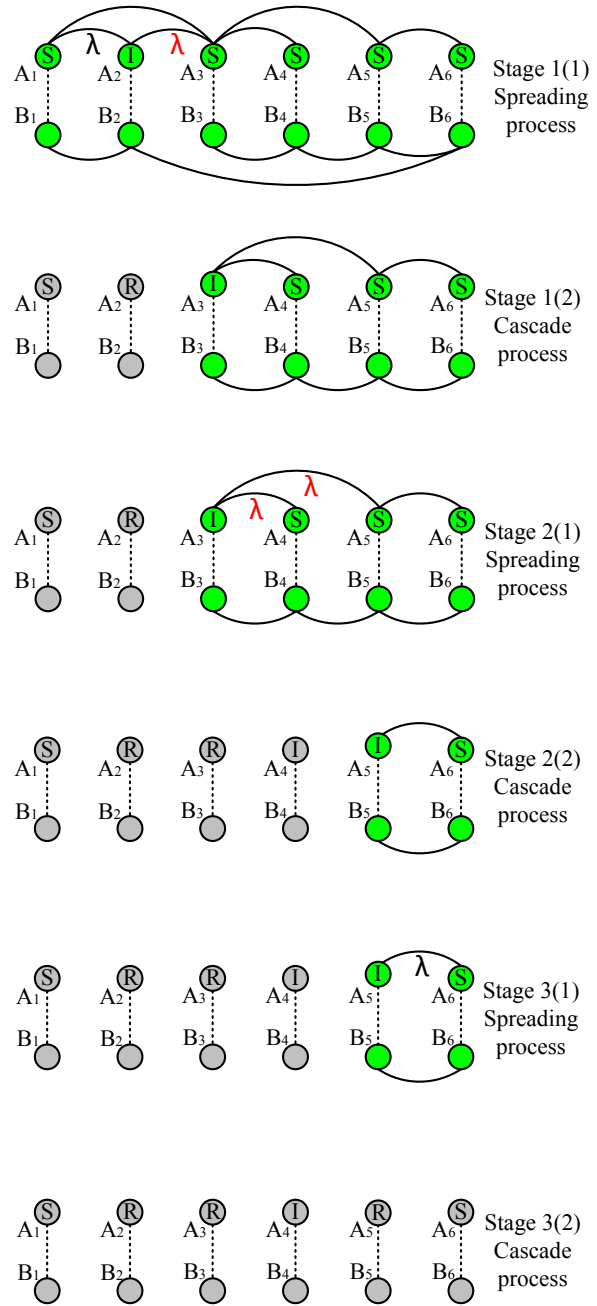} \caption{A schematic illustration of CF-VP model in interdependent networks.
Green nodes represent the functional nodes, while grey nodes
represent the removed nodes. Each stage is composed of two substages
(see the number in the brackets): disease spreading process and
cascade process. Besides, red $\lambda$ means the successful propagation of
virus, while black is the opposite case.} \label{fig.1}
\end{figure}

\section{Model}

Before defining the detailed model, we first survey the general cascading failures of interdependent networks and the SIR dynamics which we use as the paradigmatic example for the collapse process of interdependent systems under attack of the spread of virus.

The general cascading failures in interdependent networks were first proposed in Ref. \cite{buldyrev2010catastrophic}, where there are two networks A and B with the same size of $N$ nodes, then both of them are coupled via one-to-one interdependence. If node A$_i$ (B$_i$, $i = 1,2,\ldots,N$) stops function owing to attack or failure, its inter-layer counterpart B$_i$ (A$_i$) becomes nonfunctional as well. When some nodes on network A (hereafter A-nodes) are removed, the nodes of network B (hereafter B-nodes) that connect to the nonfunctional A-nodes will also be removed (because of the dependence between both networks), which further  prunes connections of these B-nodes with the giant component of network B. Subsequently, A-nodes that connect to the non  functional B-nodes will stop function and cut their connections with the new giant component of network A (only the nodes that belong to the giant component of network remain functional). These cascade processes repeat until no A-nodes and B-nodes could be removed.

\begin{figure}[!htb]
\includegraphics[scale=0.5,trim=0 0 0 0]{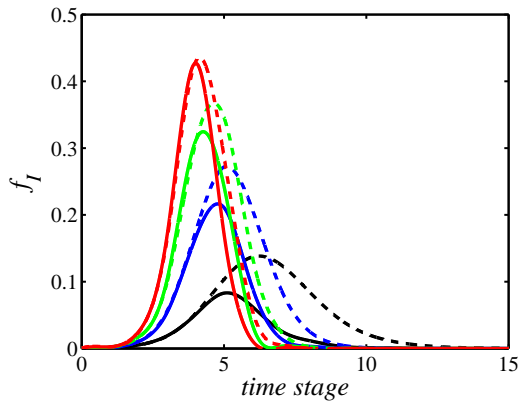} \caption{Fraction $f_I$ of infected nodes versus the time stage. The solid and dash lines represent results on network A of interdependent systems and single-layer networks, respectively. The interdependent networks are SF networks with size $N=2,000$ and same average degree $\langle k_A\rangle =\langle k_B\rangle=$ 4 (black), 6 (blue), 8 (green), and 10
(red). The transmissibility probability is $\lambda=0.5$.} \label{fig.2}
\end{figure}

SIR model \cite{pastor2015epidemic}, as one of the most fundamental and important paradigms of epidemic dynamics, classifies the network nodes into three states: susceptible (S), infected (I), or recovered or removed (R). Susceptible nodes are free of epidemic and can get infection via direct contacts with infected counterparts. Infected nodes are assumed to carry the disease and pass it towards susceptible nodes. Recovered (removed) state means the nodes recovered (died) from the disease so that these nodes neither diffuse the infection nor be infected again. In addition, classic SIR model considers discrete time process: at each time step, the infected node can infect its susceptible neighbors with transmission rate $\lambda$, and then becomes recovered or removed state with recovery rate $\delta$.

Now, we turn to our model: cascading failure and disease spreading are coupled via interdependent networks, namely, the interplay between cascading failures and virus propagation (CF-VP for short) in interdependent networks. Given the same interdependent networks as  Ref. \cite{buldyrev2010catastrophic}, we give two additional hypotheses: 1) only one network (e.g. network A) supports the propagation of virus; 2) the time scale of cascading failures is much smaller than that of virus propagation, so that virus propagation can repeat until there is no failure node in the systems. Moreover, our model also considers discrete time-step: each time stage contains the virus propagation process and one general cascading failure process. Initially, one random chosen A-node is infected by the virus, then the infected node propagates the virus: it infects its susceptible neighbors with probability $\lambda$, and then becomes removed state with probability $\delta$ (without loss of generality, we use $\delta$ = 1). In particular, if the removed nodes are assumed to be nonfunctional, a general cascading failure process will be triggered in interdependent networks and more nodes may be pruned. If there still exist infected nodes in the networks after the cascading process, a new virus propagation process and the subsequent triggered cascading failure will repeat until no infected nodes exist in the network.

To get a better understanding, Fig.~\ref{fig.1} provides a schematic example for this novel CF-VP model. Assume A$_2$ being initially infected, it can infect its neighbors with probability $\lambda$. After the spreading process of stage 1, A$_2$ and A$_3$ become removed node and infected node respectively. Due to interdependence, node B$_2$ will be nonfunctional, which subsequently causes node B$_1$ to be removed since it does not belong to the giant component of network B. Similarly, A$_1$ is removed because of the removal of B$_1$. The first stage ends. Now there exists a new infected node A$_3$, it can bring infection to its neighbors A$_4$ and A$_5$. Since A$_3$ becomes nonfunctional soon, new cascade process is triggered: B$_3$ is removed due to losing dependent counterpart; A$_4$ is removed because of separation from the giant component of network A, which in turn causes B$_4$ nonfunctional. In stage 3, even if A$_5$ fails to infect its neighbor, itself and its partner B$_5$ will also be nonfunctional due to the state transition I$\rightarrow$R of A$_5$. Because no giant component exists, A$_6$ and B$_6$ are finally removed and the systems are completely collapsed. From this illustration, it is clear that the virus propagation causes cascading failures, while the cascading failures suppress the virus propagation: S-state node A$_1$ and I-state node A$_4$ are isolated owing to the cascading failures.

\section{Results}

\begin{figure}[!htb]
\includegraphics[scale=0.5,trim=0 0 0 0]{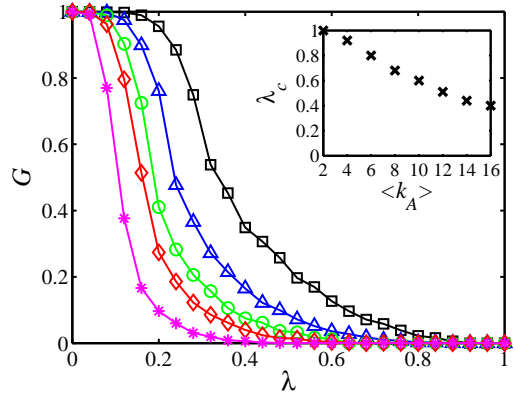}\caption{The size $G$ of
the remaining giant component of network A versus the transmissibility probability
$\lambda$. The interdependent networks are SF networks with average
degree $\langle k_B\rangle =8$, $\langle k_A\rangle=$ 4 (squares), 6 (triangles),
8 (circles), 10 (diamonds), and 16 (stars), respectively. The inset features how the threshold $\lambda_c$ changes as a function of $\langle k_A\rangle$.}
\label{fig.3}
\end{figure}

Results of computation simulations are obtained on interdependent scale-free (SF) networks with average degree $\langle k_A\rangle$ and $\langle k_B\rangle$ of networks A and B. In each CF-VP process, we assume that, initially, only one randomly chosen node is infected on network A. What we are interested is the robustness of interdependent networks against CF-VP process, which is measured by the size $G$ of the remaining giant component of network A when CF-VP ends. Here, it is worth mentioning that CF-VP model finally generates the identical size of the remaining giant component on networks A and B.

\begin{figure}[!htb]
\includegraphics[scale=0.5,trim=0 0 0 0]{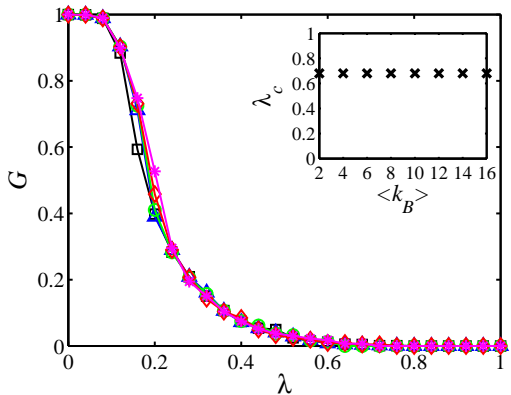}\caption{The size $G$ of
the remaining giant component of network A versus transmissibility probability
$\lambda$. The interdependent networks are SF networks with average
degree $\langle k_A\rangle =8$, $\langle k_B\rangle=$ 4 (squares), 6 (triangles),
8 (circles), 10 (diamonds), and 16 (stars), respectively. The inset features how the threshold $\lambda_c$ changes as a function of $\langle k_B\rangle$.}
\label{fig.4}
\end{figure}

Fig.~\ref{fig.2} shows the evolution of virus on network A with the proposed CF-VP
model, which is featured by the solid lines. To take a direct comparison, we also
add the traditional case of virus propagation on single-layer networks (dash lines, i.e.
without the interplay of cascading failure and virus propagation). It is obvious that though the fraction of infected nodes in both scenarios is almost identical at the early
stages, the following trend becomes greatly different. Comparing with traditional case, CF-VP model not only makes infection reach an peak faster, but also impedes the total infection risk. In fact, it is easy to elucidate these phenomena.  At the early stages, only a small fraction of nodes are infected and removed, the interdependent networks are not broken and most nodes are still functional. Thus, the virus propagates on network A almost as on single-layer networks. But with continual propagation of virus, the triggered cascading
failures cause more nodes be removed and make interdependent networks collapse into the
unconnected fragments. In particular, many infected and susceptible
nodes are also removed due to the cascading failures, which in turn leads
to the effective suppression of virus propagation (also see Fig.~\ref{fig.1}). With the CF-VP framework, the role of feedback loop becomes clear: virus propagation induces cascading failure, while cascading failure suppresses virus propagation.

Besides, another interesting observation from Fig.~\ref{fig.2} is that, similar to traditional case, the spreading scale of virus is larger in network A with denser connections (i.e. the larger the average degree, more obvious the infection peak will be), which makes the total transmission become easier. Due to feedback loop (refer to Fig.~\ref{fig.1}), this should in turn cause larger-scale cascading failures in interdependent systems and make systems more vulnerable to CF-VP model, which we will systematically discuss in what follows.

To explore the influence of CF-VP model on the robustness of interdependent networks, we focus on two opposite cases. The first case is to fix the average degree of network B ($\langle k_B\rangle$) yet vary the average degree of network A ($\langle k_A\rangle$); another case is to fix $\langle k_A\rangle$ yet vary $\langle k_B\rangle$ (Indeed, there exists the third case: keep $\langle k_A\rangle$ and $\langle k_B\rangle$ equal, i.e. $\langle k_A\rangle=\langle k_B\rangle$, and simultaneous changing, like Fig.~\ref{fig.2}. But here we do not plot the curves of this case, which will be explained soon). Interestingly, such a change that seems trivial will lead to greatly different outcomes. First, irrespective of which case, increasing $\lambda$ makes $G$ become smaller, namely, fast propagation of virus will trigger larger, stronger crash of systems. In particular, there exists the critical threshold of virus transmissibility, $\lambda_c$, above which the remaining giant component will be null. From Fig.~\ref{fig.3}, we can see that $\lambda_c$ becomes smaller with the increment of $\langle k_A\rangle$, which means that denser connections of network A where virus spreading takes place will accelerate the propagation of virus and thus the progress of cascading failures in systems. At variance, the change of $\langle k_B\rangle$ has no obvious impact on the threshold $\lambda_c$ (see Fig.~\ref{fig.4}), i.e. the crash trend is nearly identical if only $\langle k_B\rangle$ changes. Combining these observations, a significant finding poses itself: the interdependent systems will be more vulnerable only if the network layer supporting virus propagation has denser connections (i.e. larger average degree); otherwise the interdependent systems are robust against the change of connections in other layer(s). Along this discovery, it now becomes easy to understand that simultaneously changing $\langle k_A\rangle$ and $\langle k_B\rangle$ will generate the same results as Fig.~\ref{fig.3}. In addition, this discovery also differs from previous framework of cascading failure in interdependent networks [11], where better robustness usually needs denser connections. Thus, our outcomes, to some extent, prove the necessity and significance of feedback loop when designing the interdependent networks.

\section{Control Strategy}

Up to now, it has been very clear that in interdependent networks virus propagation on one layer could lead to continuous cascading failures and fragmentation of systems. Along this line, the most intuitive method of protecting interdependent networks is to control the spread of virus when it appears. In reality, it seems hard to timely restrain the spreading of virus (especially the emerging virus) by using the well-known pre-immunization strategies \cite{pastor2002immunization,cohen2003efficient}, due to the absence of effective antivirus programs \cite{gao2013modeling}.
However, in CF-VP model it seem feasible to identify the infected neighbor based on knowledge and abnormal behavior of infected nodes. Here we consider such a control strategy: after the emergence of virus, susceptible node $i$ can identify one infected neighbor with probability $q_i$, and then prunes its connection with this neighbor. This strategy not only isolates the health nodes from their infected neighbors, but also
decreases the average degree of network layer which supports the virus propagation (see Fig.~\ref{fig.3} for its impact). With respect to the identification capability $q_i$, we consider two following cases.

\begin{figure}[!htb]
\begin{minipage}[l]{0.48\linewidth}
\includegraphics[scale=0.31,trim=0 0 0 0]{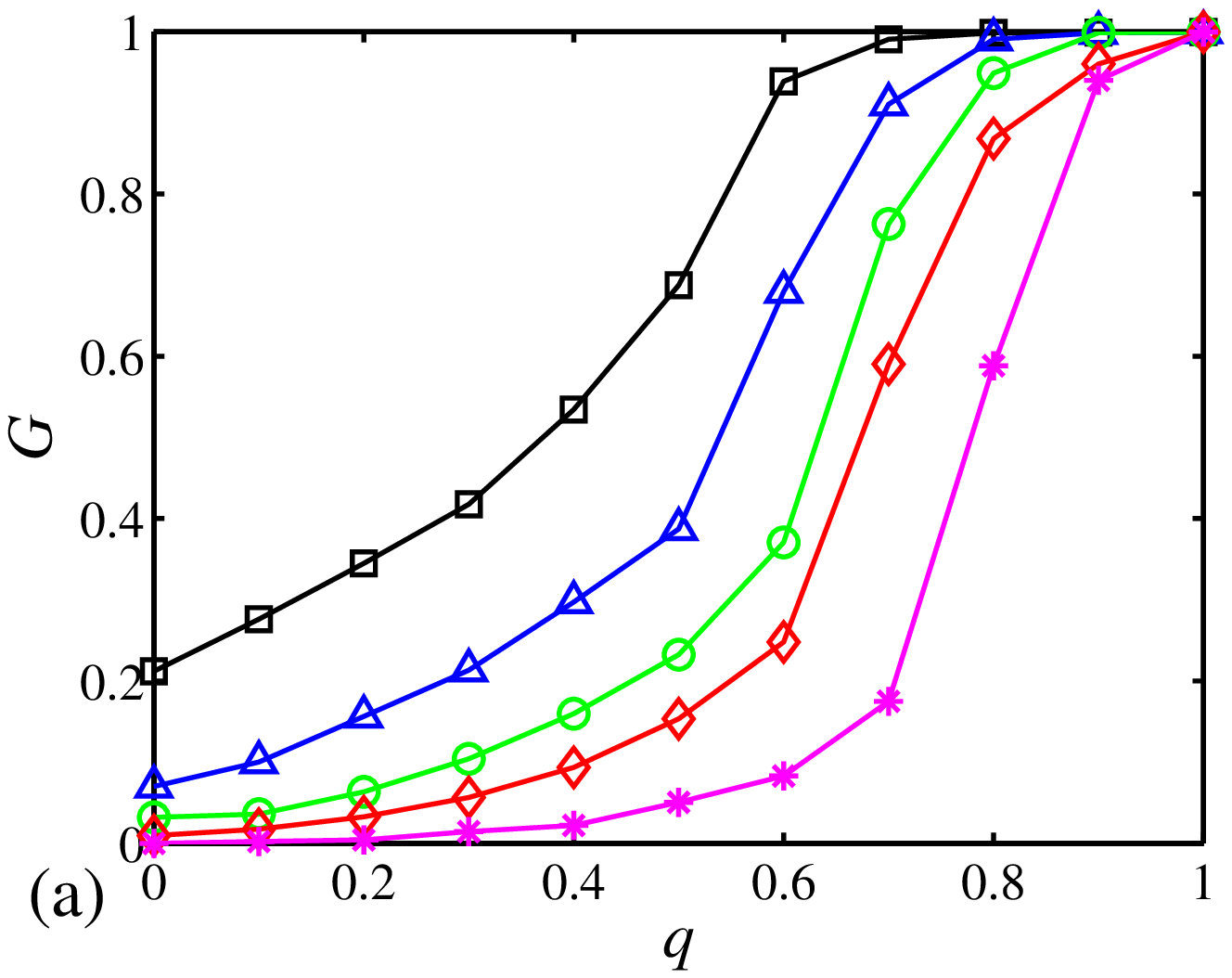}
\end{minipage}
\begin{minipage}[r]{0.5\linewidth}
\includegraphics[scale=0.31,trim=0 0 0 0]{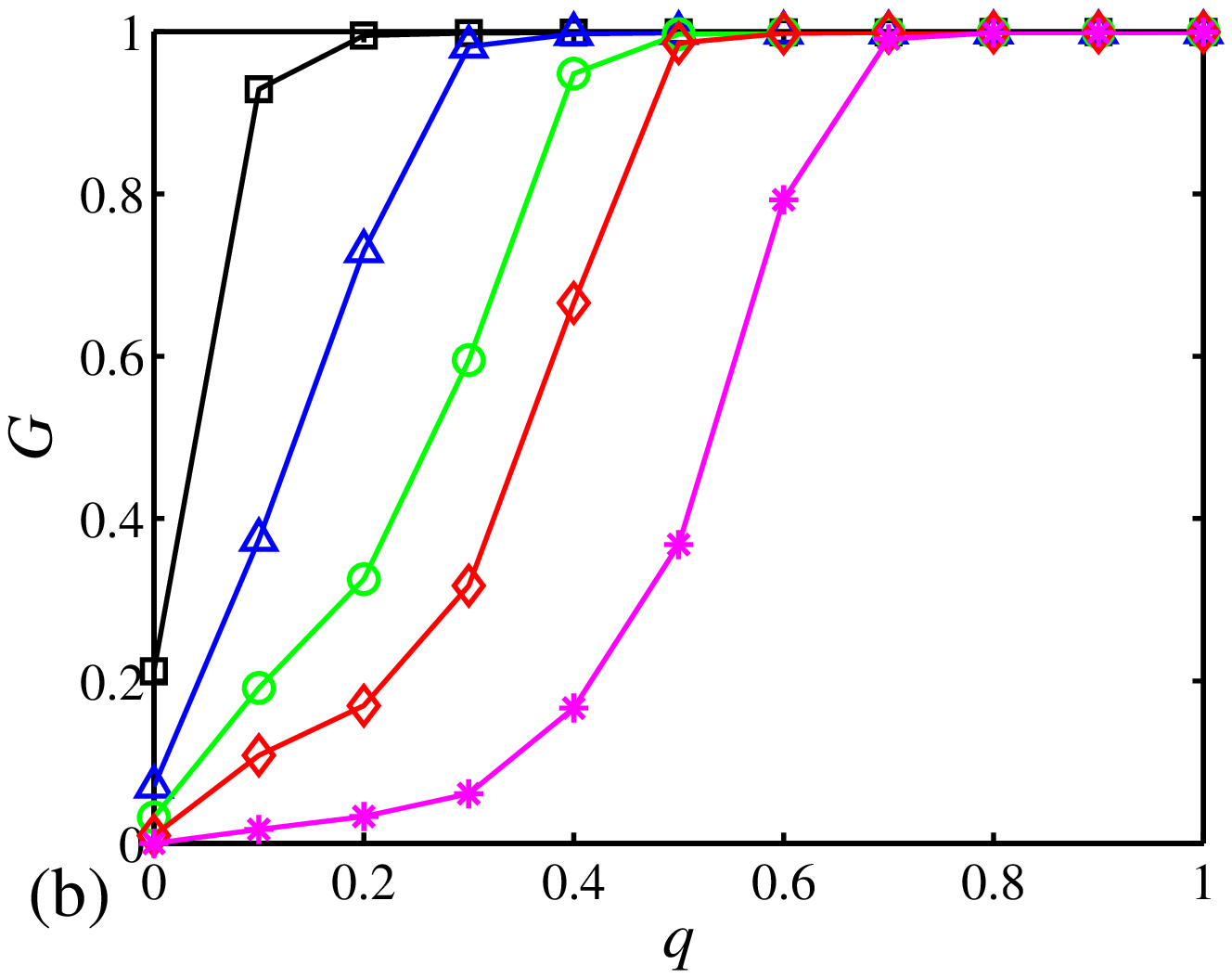}
\end{minipage}
\caption{The size $G$ of the remaining giant component of network A versus identification
probability $q$ for deterministic adaptive isolation case (a) and degree-based adaptive isolation where $\sigma=0.3$ (b). The interdependent networks are SF networks
with average degree $\langle k_B\rangle = 8$, $\langle k_A\rangle=$ 4 (squares), 6 (triangles), 8 (circles), 10 (diamonds) and 16 (stars), respectively. The
transmissibility probability is $\lambda=0.5$.}
\label{fig.5}
\end{figure}

1) Deterministic adaptive isolation: $q_i=q_j=q$ for $i\neq j$. That is, all of nodes have
the same ability to idenfify infected neighbors.

2) Degree-based adaptive isolation: $\{q_1,q_2,\cdots, q_N\}$ following gauss distribution. That is, $\{q_1,q_2,\cdots, q_N\}\sim N(q, \sigma)$, where $q$ and $\sigma$ are mean and
standard deviation respectively. Moreover, if $k_i \geq k_j$, we assume $q_i\geq q_j$, which means that large-degree nodes have higher ability to identify infected neighbors. Considering that $q_i$ must be between 0 and 1, we assign $q_i$ as

\begin{equation}q_i=
\left\{\begin{array}{ll} 0,\ \emph{\emph{if}}\ \ q_i<0,\\
q_i,\ \ \emph{\emph{if}}\ \ 0\leq q_i\leq 1\\ 1,\ \ \emph{\emph{if}}\ q_i>1,\end{array}\right.
\end{equation}

Subsequently, we explore how the control measures improve the robustness of interdependent
networks under CF-VP model, where we still use two opposite cases as Figs.~\ref{fig.3} and~\ref{fig.4}. Fig.~\ref{fig.5} first shows the impact of isolation strategies when
$\langle k_B\rangle$ is fixed and $\langle k_A\rangle$ changes.  It is clear that the size $G$ of remaining giant component increases with $q$, which indicates the robustness of interdependent networks can be significantly improved by increasing nodes' identification capability, regardless of which strategy. With large identification probability, the infection source(s) can be controlled and isolated earlier. The removal of these infected nodes further makes the cascading process become slow, which also decreases the possibility of infection outbreak. This thus validates the importance of feedback loop in the coupled disease-cascading model once again. Moreover, another similar phenomenon in Figs.~\ref{fig.5} (a) and (b) is that network A possessing large average degree needs larger $q$ to maintain the equivalent robustness with the case of small $\langle k_A\rangle$, which in fact is consistent with the prediction of Fig.~\ref{fig.3}: larger $\langle k_A\rangle$ usually enables systems to become more vulnerable, thus requiring more powerful protection. Except for similarity, we can also notice that degree-based adaptive isolation performs much better than deterministic adaptive isolation. This actually agrees with our intuition,  because (as single-layer networks) large-degree nodes play a more significant role in the propagation of virus than small-degree nodes. If there exist infected nodes among the neighborhood of large-degree nodes, they can easily prune the connections with infected neighbor(s) due to large identification ability. With fast removal of infection sources, cascading will be controlled better (i.e. larger $G$ for the same $q$ value).

\begin{figure}[!htb]
\begin{minipage}[l]{0.48\linewidth}
\includegraphics[scale=0.31,trim=0 0 0 0]{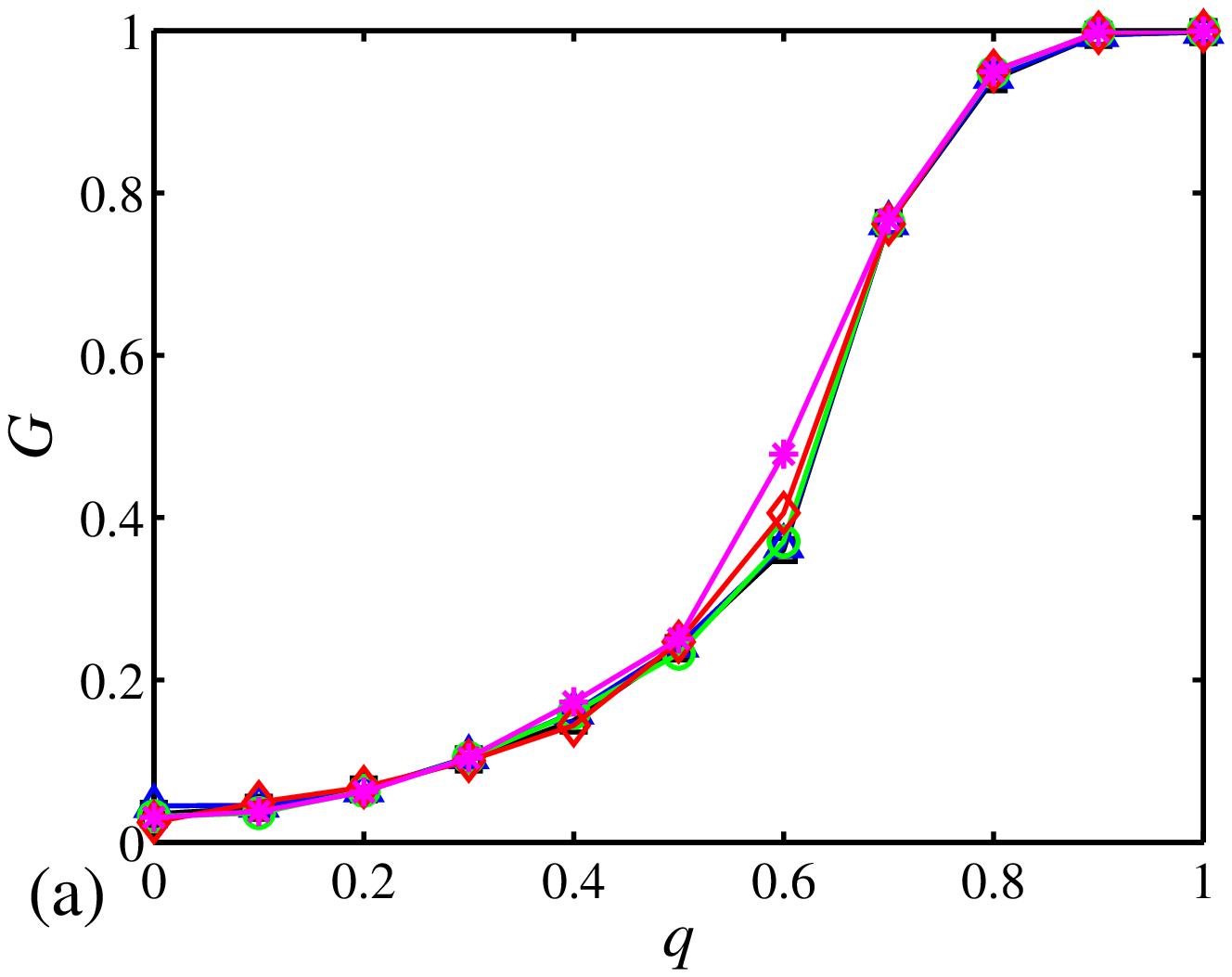}
\end{minipage}
\begin{minipage}[r]{0.5\linewidth}
\includegraphics[scale=0.31,trim=0 0 0 0]{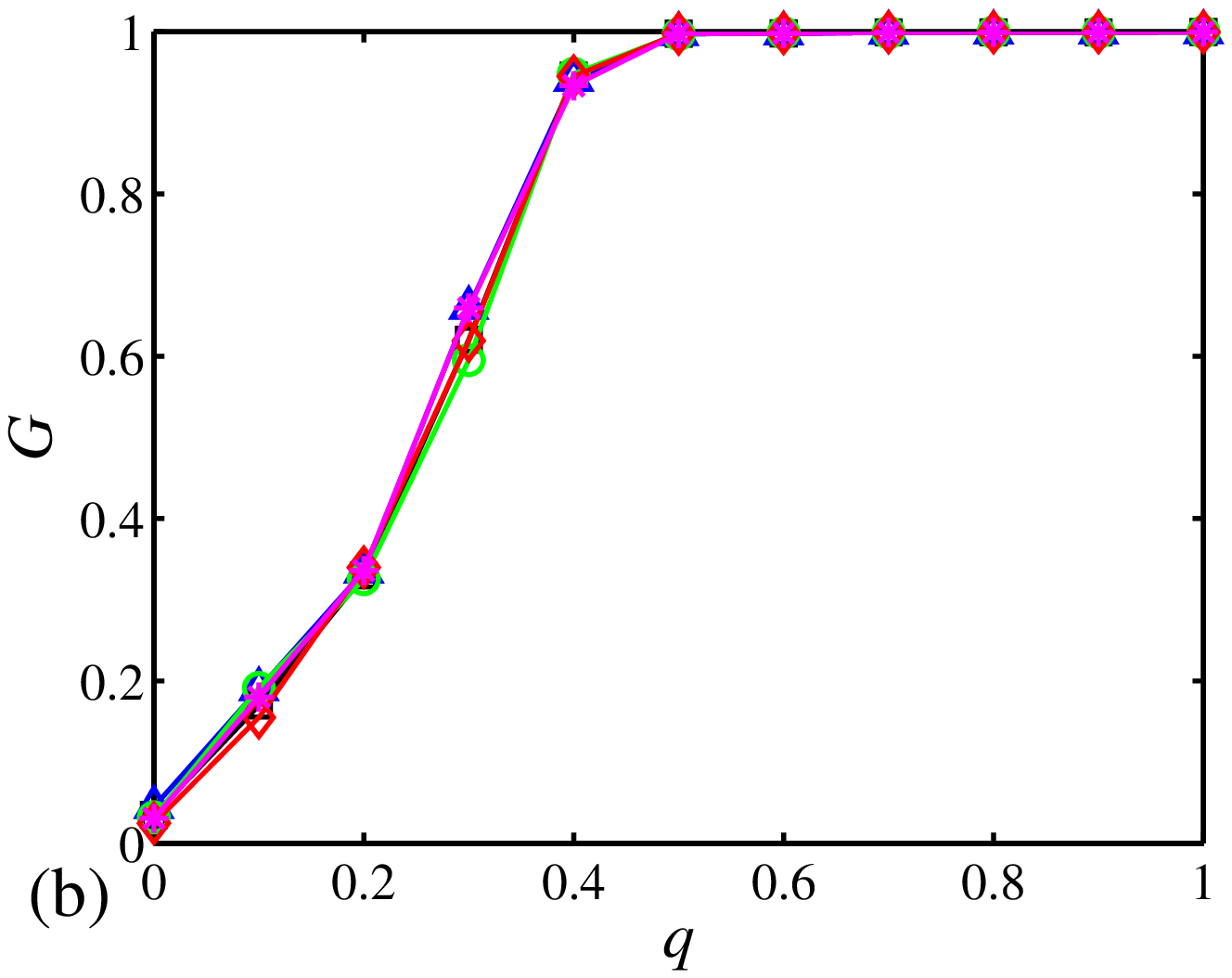}
\end{minipage}\caption{The size $G$ of the remaining giant component of network A versus identification probability $q$ for deterministic adaptive isolation case (a) and degree-based adaptive isolation where $\sigma=0.3$ (b). The interdependent networks are SF networks
with average degree $\langle k_A\rangle = 8$, $\langle k_B\rangle=$ 4 (squares), 6 (triangles), 8 (circles), 10 (diamonds) and 16 (stars), respectively. The
transmissibility probability is $\lambda=0.5$.}
\label{fig.6}
\end{figure}

We now turn to another case: fixing $\langle k_A\rangle$ yet varying $\langle k_B\rangle$ and study how the isolation strategies improve the robustness of interdependent networks. As reflected in Fig.~\ref{fig.4}, this case has no impact on the system crash. Though the size $G$ of remaining giant component enhances with identification capability $q$ and degree-based adaptive isolation performs better, only changing $\langle k_B\rangle$ will generate nearly identical results with each isolation strategy (i.e. the overlapped curves in Fig.~\ref{fig.6}). This is because, for each $q$ value, the isolation probability of infected neighbors on network supporting virus propagation is the same, irrespective of average degree in other network. Combining Figs.~\ref{fig.5} and~\ref{fig.6}, it seems to indicate that the best way of controlling system crash is to eradicate the infection sources in epidemic layer, which is specially useful for this layer with denser connections.

\section{Conclusion}

In this letter, we have developed a toy model (CF-VP model) in which virus propagation and cascading failure are coupled on interdependent networks, where each network layer sustains one dynamic process. For both processes, they spontaneously form a novel feedback loop: virus propagation triggers continuous cascading failures and even complete fragmentation if transmissibility probability is above a threshold; while cascading failures will break the connections of networks and thus suppresses virus propagation. Of note, if the network layer supporting epidemic spreading has denser connections, interdependent systems will be more vulnerable, which is opposite to the observation of traditional cascading fashion in interdependent networks \cite{buldyrev2010catastrophic}. To protect interdependent networks, we further propose the control measures based on the capability to identify the infected neighbor. Interestingly, the larger the identification capability (especially for larger-degree node), more robustness the interdependent networks will be.

In spite of simplicity, our model describing the interplay between cascading failures and virus propagation in interdependent networks seems reasonable and as well easily justifiable with realistic situations. For example, Internet and some social online networks could be encapsulated into the framework of multilayer networks. But how they influence each other will be a long-term question. This work may provide some new insights into understanding the interplay and proposing the protection measures. Besides, another point that deserves our attention is to consider theoretical analysis framework, which may validate the present simulation findings. Except for interplay between dynamical processes, the co-evolution between dynamics and interdependent network topology is also worth of our endeavors in future.

\begin{acknowledgments} This work is supported by the Shandong Province Outstanding Young Scientists Research Award Fund Project (Grant No. BS2015DX006), the Shandong Academy of Sciences Youth Fund Project (Grant No. 2016QN003), the Inner Mongolia Colleges and Universities Scientific and Technological Research Projects (Grant no. NJZY132), and the National Natural Science Foundation of China (Grant Nos. 61572297, 31560622, 31260538, 30960246).\end{acknowledgments}

\bibliographystyle{unsrt}
\bibliography{reference}
%

\end{document}